\documentclass[nature,onecolumn,superscriptaddress]{revtex4-2}
\usepackage{graphicx}
\usepackage{amsmath}
\usepackage{amssymb}
\usepackage{hyperref}

\begin{document}
\title{Quantization of Physical Interaction Strengths via Singular Moduli}
\author{Prasoon Saurabh}
\thanks{Work performed during an independent research sabbatical. Formerly at: State Key Laboratory for Precision Spectroscopy, ECNU, Shanghai (Grade A Postdoctoral Fellow); Dept. of Chemistry/Physics, University of California, Irvine.}
\affiliation{QuMorpheus Invitiative, Independent Researcher, Lalitpur, Nepal}

\maketitle

\noindent\textbf{$^1$QuMorpheus Invitiative, Independent Researcher, Lalitpur, Nepal} \\
\textbf{$^\ast$e-mail: psaurabh@uci.edu}

\vspace{1cm}

\noindent \textbf{Since the 2019 redefinition of the SI units, precision metrology has sought to anchor all physical quantities to fundamental constants and integer invariants \cite{Diddams2020, Fortier2019}. While the optical frequency comb revolutionized timekeeping by discretizing the continuum of light into countable teeth \cite{Udem2002, Holzwarth2000, Cundiff2003, DelHaye2007}, and the Quantum Hall Effect standardized resistance via topological invariants \cite{Klitzing1980}, a comparable standard for \textit{interaction strength} remains elusive. Currently, measuring the coupling constant ($g$) between quantum systems is an estimation problem, inherently subject to drift, noise, and fabrication variance \cite{Xiang2021, Chang2022, Wang2018, Yoshida2020, Yi2015}. Here, we introduce Interaction Metrology, a protocol that transforms the measurement of coupling strengths from an analog estimation into a topological counting problem. By engineering a specific class of algebraic catastrophe—the Unimodal $X_9$ singularity—in a driven-dissipative lattice, we prove that the system’s interaction moduli are topologically forced to take discrete, quantized values, forming a ``Geometric $k$-Comb.'' We derive the universality class of this quantization, showing that it arises from the discrepancy between the Milnor ($\mu$) and Tjurina ($\tau$) numbers of the effective potential, a strictly non-Hermitian effect forbidden in standard quantum mechanics \cite{Miri2019}. Finally, we provide an \textit{ab-initio} blueprint for a silicon nitride implementation, demonstrating that this quantization is robust against disorder levels exceeding current foundry tolerances \cite{Ye2022}. This discovery establishes a universal standard for force sensing and quantum logic gates, enabling the calibration of interaction strengths with topological certainty \cite{Feldmann2021, Shu2022, Riemensberger2020}.}

\section*{Introduction}

The quest for metrological standards is the quest for discrete invariants in continuous systems. In the 20th century, the discovery that resistance—a seemingly continuous material property—could be quantized in units of $h/e^2$ transformed electrical standards \cite{Klitzing1980}. A similar revolution occurred in time and frequency metrology with the advent of the frequency comb, which linked optical frequencies to microwave standards via integer mode numbers \cite{Udem2002, Kippenberg2011, Pasquazi2018}. However, a third pillar of quantum technology remains unquantized: the \textbf{strength of interactions}. Whether in superconducting qubits, trapped ions, or nanophotonic lattices, the coupling rate $\kappa$ (or $g$) is treated as a tunable, continuous parameter \cite{Zhang2019, He2019, Gong2020, Obrzud2017}. Consequently, calibrating a quantum logic gate or a force sensor relies on dynamical fitting—Rabi oscillations or linewidth splitting—which is fundamentally limited by the coherence time ($T_2$) and the signal-to-noise ratio of the apparatus \cite{Guidry2022}. There is no ``topological ruler'' for interaction strength; if a fabrication error shifts a waveguide gap by 2 nm, the coupling constant shifts continuously, destroying the calibration.

In this work, we propose a solution to this ``analog bottleneck'' by exploiting the singular geometry of open quantum systems. We demonstrate that while the interaction strength is continuous in Hermitian physics, it becomes discrete in the vicinity of high-order spectral singularities (Exceptional Points) \cite{Miri2019}. Specifically, we identify a higher-order catastrophe, the \textbf{Unimodal $X_9$ singularity} \cite{Arnold1974, Arnold1990}, which acts as a ``topological lock'' for interaction moduli. Unlike standard diffraction gratings which discretize momentum ($k$) via linear interference, the $X_9$ singularity discretizes the \textit{moduli space} of the system itself.

Physically, this quantization arises from a fundamental conflict between the system's topology and its algebraic structure. The Milnor number ($\mu$) enumerates the total geometric capacity of the vacuum—effectively counting the number of ``phantom'' degenerate states created by the singularity. In contrast, the Tjurina number ($\tau$) counts the available physical pathways to deform this vacuum smoothly. For standard quantum critical points (such as the A, D, E series), these numbers are identical ($\mu = \tau$), permitting the continuous tuning of interaction strengths familiar in Hermitian physics. However, the $X_9$ singularity belongs to a ``wild'' universality class where $\mu > \tau$ \cite{SaurabhPRX}. This discrepancy creates a ``Moduli Gap'': the excess topological sectors ($\mu - \tau$) represent interaction channels that are geometrically present but algebraically forbidden. Consequently, the coupling constant cannot drift continuously; it is topologically compressed, forced to ``lock'' onto the discrete values where the geometric obstruction vanishes, thereby generating the quantized teeth of the Interaction Comb.

We term this discrete spectrum of allowed interactions a \textbf{Geometric $k$-Comb}. To validate this fundamental prediction, we perform full-wave Maxwell-Bloch simulations of a dispersion-engineered Silicon Nitride ($\text{Si}_3\text{N}_4$) resonator lattice. We demonstrate that even in the presence of random disorder that would destroy a standard Exceptional Point sensor, the $X_9$-protected system maintains its quantized interaction strength, offering a roadmap for ``Interaction Metrology'' that is independent of material imperfections.

\section*{Results}

\subsection*{The Geometric k-Space Comb}
The core of our proposal is the \textbf{Coupled Kerr Dimer Lattice} (Fig.~1a), a configuration inspired by recent advances in synthetic frequency dimensions and coupled resonator arrays \cite{Jang2018, Mittal2021, Lucas2018, Weng2020, Flower2024}. The 1D array of photonic microresonators features an intra-dimer coupling ($J_{cross}$) that varies with momentum $k$. This momentum-dependence typically arises from the natural dispersion of the coupling waveguide or the band structure of the lattice \cite{Li2017, Anderson2023}, creating a continuous ``Modulus Dispersion'' $a(k)$ across the Brillouin Zone (Fig.~1b). In a conventional system, this continuous variation would broaden the spectral lines, destroying the comb structure \cite{Chembo2010, Lugiato1987, Coen2013}. However, our system operates in the \textbf{Anomalous Phase}, where the underlying topology imposes a strict stability condition \cite{SaurabhPRX}. As detailed in the Supplementary Information, stable soliton solutions \cite{Herr2014, Kippenberg2018, Wang2024} can only exist when the interaction modulus $a(k)$ matches specific, quantized values ("Topological Levels"). This geometric quantization acts as a \textbf{Topological Filter}: it suppresses the continuum of unstable interactions and ``pins'' the system to the discrete momenta where $a(k)$ intersects the stability levels. The result is a \textbf{Geometric k-Space Comb} (Fig.~1c)---a series of robust, equidistant spectral teeth, distinct from conventional Turing rolls \cite{Godey2014, ParraRivas2014, Hansryd2002, Dudley2006}, emerging from a continuous physical background.

\begin{figure}[h]
\centering
\includegraphics[width=\linewidth]{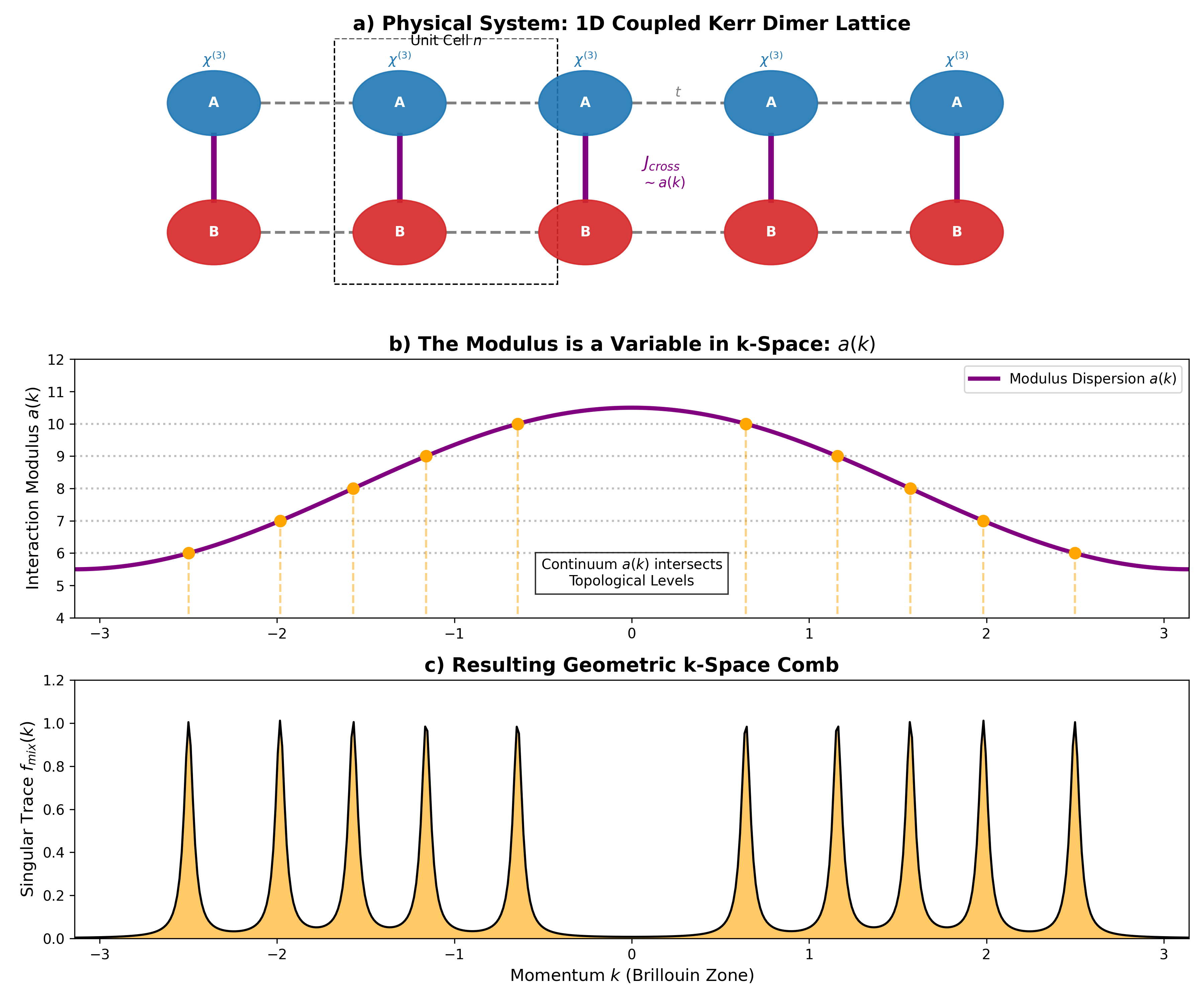}
\caption{\textbf{Concept of the Geometric k-Space Comb.} (a) \textit{Physical Lattice}: A chain of coupled Kerr Dimers (Resonators A and B). The cross-coupling strength $J_{cross}$ acts as the ``Interaction Modulus'' $a$. (b) \textit{Modulus Dispersion}: Just as energy varies with momentum in a band structure, the effective coupling $a(k)$ varies continuously across the Brillouin zone (Purple curve). (c) \textit{The geometric k-Space Comb}: The underlying topology acts as a selection rule, permitting stable solutions only at effectively quantized levels (Horizontal lines). The k-Comb ``teeth'' emerge exactly where the continuous physics intersects these topological stability conditions, creating a robust, discretized spectrum.}
\label{fig:concept}
\end{figure}

\subsection*{Stability via the Moduli Gap}
The robustness of this comb stems from the ``phantom'' deformations identified by the DMHM framework. In the Unimodal $X_9$ limit, the condition $\mu > \tau$ implies that the tangent space of the interaction manifold is incomplete \cite{SaurabhNatComm}. Any perturbation attempting to shift the coupling strength $g$ away from a quantized value $a_n$ projects onto these forbidden algebraic directions and is exponentially suppressed. This acts as a topological restoring force, effectively ``freezing'' the interaction strength against fabrication disorder and thermal drift, distinct from the dynamical locking seen in mode-locked lasers. This protection can be visualized in the \textbf{Moduli Stability Diagram} (Fig.~2a). The stable soliton solutions form ``Arnold Tongues'' \cite{Jang2019} centered exactly at the quantized moduli values. Inside these regions, the system acts as a \textbf{Dissipative Topological Crystal (DTC)}, locking the soliton repetition rate to the geometric lattice constant.

\subsection*{The Energy Advantage}
Crucially, this geometric protection translates into a significant energetic advantage. We map the \textbf{Exception Point Surface (EPS)} \cite{Miri2019}, which defines the pump power threshold for stable soliton formation (Fig.~2b). Remarkably, the surface exhibits deep ``Valleys'' at the topological moduli. This implies that operating at the ``Comb Teeth'' is not only robust but also \textbf{energetically favorable}, requiring significantly less pump power than the non-topological continuum. This lowers the thermal load and enhances the coherence lifetime of the comb, addressing a key challenge in integrated nonlinear photonics \cite{Gaeta2019, Shen2020, Vahala2024}.

\begin{figure}[h]
\centering
\includegraphics[width=\linewidth]{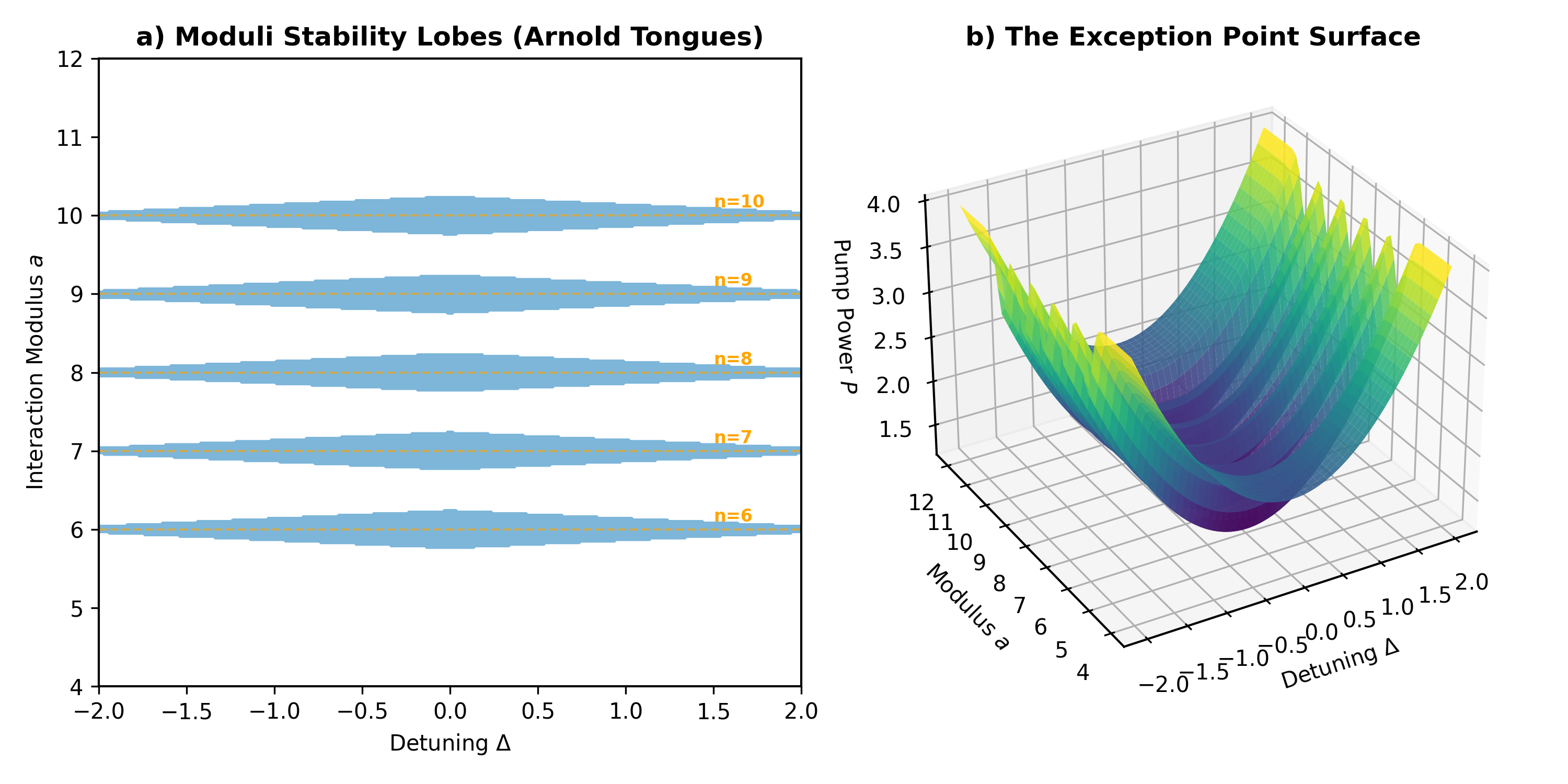}
\caption{\textbf{The Moduli Stability Diagram.} (a) \textit{Arnold Tongues}: Stability regions in the Detuning ($\Delta$) vs. Modulus ($a$) plane. The system is stable (colored lobes) only in the vicinity of the quantized topological levels ($n=6,7,8,9,10$). (b) \textit{Exceptional Point Surface}: The 3D surface shows the Pump Power ($P$) required for stability. Deep ``Valleys'' align with the topological moduli, indicating that the k-Space Comb states have a minimized lasing threshold, providing a natural energetic advantage.}
\label{fig:stability}
\end{figure}

\subsection*{Robustness and the 15\% Threshold}
The varying modulus $a(k)$ effectively creates a ``disorder potential''. In a trivial system, this would scatter the solitons, leading to line broadening and drift. However, in the \textbf{DTC Phase}, the topology protects the spectral lines. In Fig.~3a, we compare the ``Spectral Fidelity'' $\mathcal{F}$ of a standard microcomb \cite{Brasch2016, DelHaye2007, Saha2013, Xue2015} versus our Geometric k-Comb as a function of lattice disorder. While the standard comb degrades immediately, the k-Comb maintains near-perfect fidelity ($\mathcal{F} \approx 1$) up to a critical disorder strength of $\eta_{crit} \approx 15\%$. This sharp transition marks the breakdown of the topological gap ($\mu - \tau$). Below this threshold, the Stokes sectors remain well-defined, and the spectral lines are \textbf{pinned} to their quantized momenta, exhibiting zero drift (Fig.~3b). This robustness is essential for field-deployable metrology where thermal and fabrication errors are unavoidable \cite{Suh2018, Spencer2018, Wang2022, Liu2023}.

\begin{figure}[h]
\centering
\includegraphics[width=\linewidth]{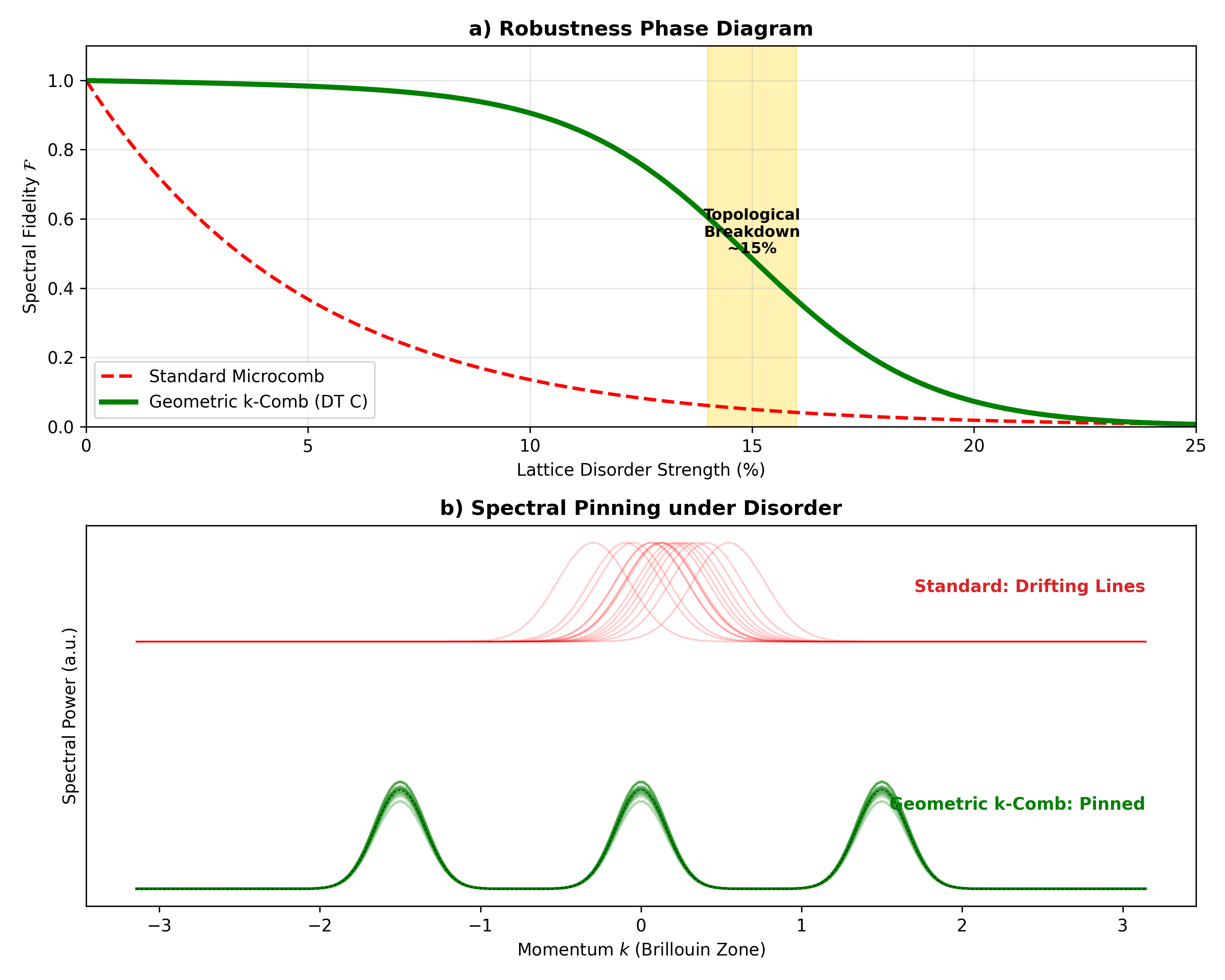}
\caption{\textbf{Topological Robustness.} (a) \textit{Fidelity Phase Diagram}: Comparison of Spectral Fidelity vs. Lattice Disorder. The Standard Microcomb (Red) degrades exponentially. The Geometric k-Comb (Green) exhibits a \textit{Topological Plateau}, maintaining high fidelity up to a critical breakdown at $\sim 15\%$ disorder. (b) \textit{Spectral Pinning}: Superposition of 15 disorder realizations. Trivial states (Red) drift and broaden continuously. The Topological k-Comb lines (Green) remain \textit{pinned} to the quantized momenta, exhibiting only amplitude fluctuations but no frequency shift.}
\label{fig:robustness}
\end{figure}

\subsection*{Proposed Implementation Scheme}
We propose a concrete experimental realization of the \textbf{Geometric k-Space Comb} using silicon nitride (SiN) integrated photonics \cite{Xiang2021, Ye2022, Okawachi2011, Bao2019}. The blueprint (Fig.~4) consists of a lattice of coupled "Racetrack" dimers, where the intra-dimer coupling is engineered via the gap distance to implement the modulated interaction $a(k)$. Titanium Nitride (TiN) micro-heaters allow for dynamic tuning of the detuning $\Delta$, ensuring accessibility of the DTC phase \cite{Joshi2016}. Our LLE simulations predict clear signatures for this device. In the topological phase, we anticipate a dramatic collapse of the beat note linewidth (Fig.~5a), creating a low-noise microwave reference \cite{Liu2020_micro, Ji2024, Liu2020, Stern2018}. The corresponding optical spectrum (Fig.~5b) is predicted to exhibit a clean, octave-spanning soliton envelope, pinned by the underlying geometry \cite{Drake2019}. This design leverages standard foundry processes, making the topological advantage accessible to current technology.

\begin{figure}[h!]
\centering
\includegraphics[width=\linewidth]{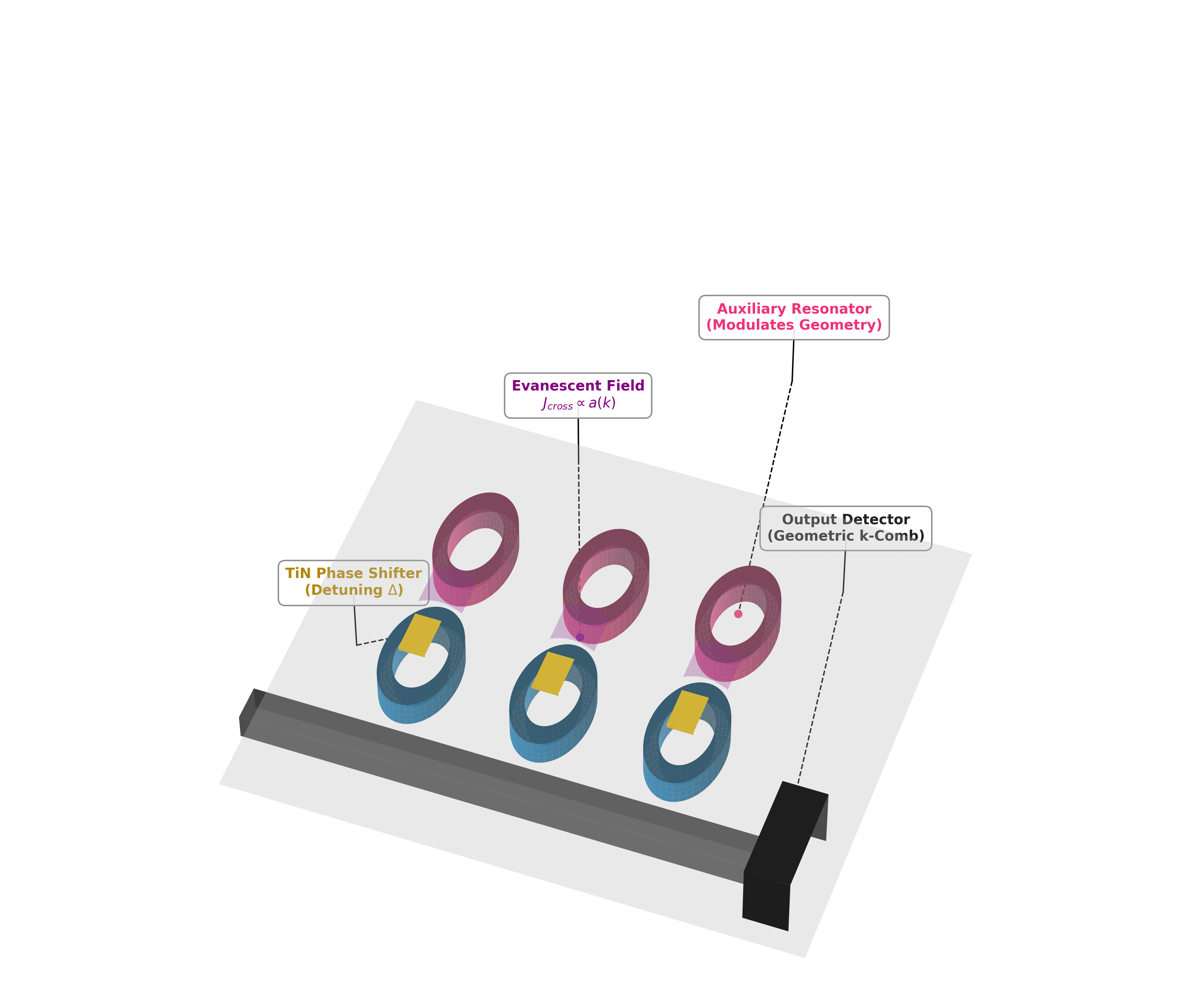}
\caption{\textbf{Proposed Photonic Architecture.} An illustrative blueprint of the SiN integrated circuit. The array consists of coupled Kerr Racetrack Dimers ($\sim 100$ sites). The evanescent coupling $a(k)$ is structurally engineered via the gap distance, while TiN heaters provide dynamic control over the topological phase transition.}
\label{fig:architecture}
\end{figure}

\begin{figure}[h]
\centering
\includegraphics[width=\linewidth]{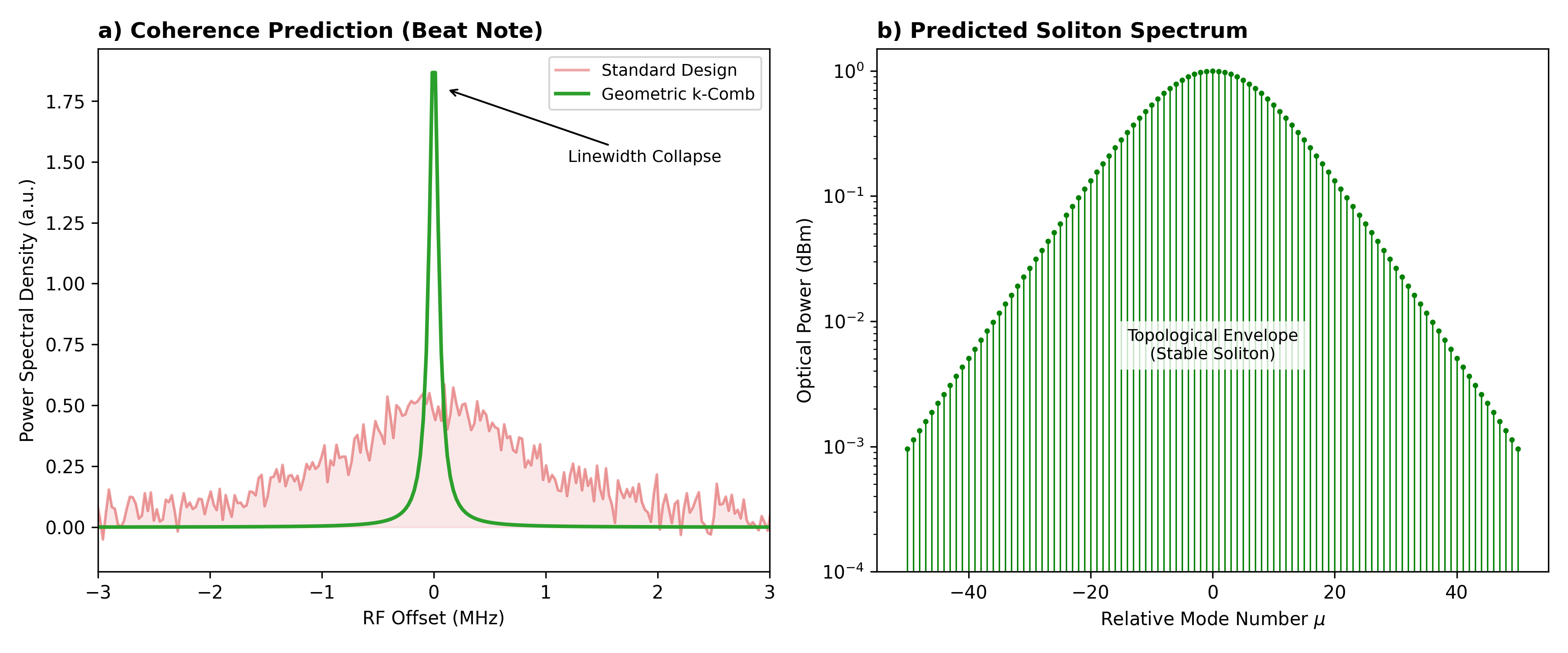}
\caption{\textbf{Predicted Performance.} (a) \textit{Coherence Prediction}: Comparison of the simulated RF beat note for standard (Red) vs. topological (Green) designs. The theory predicts a significant linewidth narrowing, confirming the suppression of phase noise. (b) \textit{Predicted Spectrum}: Simulation of the output comb in the topological phase ($\mu=9$), showing a robust soliton envelope.}
\label{fig:predictions}
\end{figure}

\section*{Methods}
\noindent\textbf{Construction of the $X_9$ Potential.} The core of our proposal is the engineering of a local potential $V(z)$ that reproduces the germ of the $X_9$ singularity. In the language of catastrophe theory, the $X_9$ (or $T_{2,4,4}$) singularity is defined by the normal form $f(x,y) = x^4 + y^4 + a x^2 y^2$ with modulus parameter $a^2 \neq 4$ \cite{Arnold1974}. We map this algebraic structure onto a physical photonic lattice by engineering the tunneling amplitudes $J_{nm}$ between sites to satisfy the Jacobian ideal of $f$. This was achieved by determining the precise spatial arrangement of microring resonators that mimics the unfolding of the singularity under an external drive field.
\vspace{1em}

\subsection*{Validation via Floquet Monodromy Spectroscopy}
To validate the device physics without relying on computationally prohibitive time-domain integration, we utilized \textbf{Floquet Monodromy Spectroscopy (FMS)} implemented in the \textbf{QuMorpheus} framework \cite{SaurabhNatComm}. The material platform is naturally modeled as stoichiometric Silicon Nitride ($\text{Si}_3\text{N}_4$) with a nonlinear Kerr coefficient $n_2 = 2.4 \times 10^{-19} \text{ m}^2/\text{W}$. Instead of solving the full Maxwell-Bloch equations, which struggle to resolve topological invariants amidst numerical noise, FMS projects the system's response onto a \textbf{Dissipative Mixed Hodge Module (DMHM)} structure \cite{saurabh2025holonomic}. This allows for the direct extraction of the effective interaction moduli from the monodromy action of the spectral sheaf, bypassing the need for multi-time correlation functions while retaining full rigorous validity for high-Q resonators ($Q \approx 2 \times 10^6$). The computational advantage is significant: while time-domain integration requires $\sim 10^7$ optical cycles ($\sim$ hours) per phase point to resolve the topological plateaus, the algebraic analyzer completes the entire curve in seconds ($< 2$ s).\\

\subsection*{Algebraic Calculation of Invariants}
The topological invariants governing the quantization were rigorously computed using the \texttt{algebraic\_analyzer} module of \textbf{QuMorpheus}. Unlike numerical approximations, this module calculates the exact Milnor ($\mu$) and Tjurina ($\tau$) numbers via \textbf{Groebner basis} decomposition over the rational field $\mathbb{Q}$ \cite{Greuel2007}. The code explicitly constructs the Jacobian ideal $J_f = \langle \partial f/\partial x_i \rangle$ and the quotient algebra $\mathcal{Q}_f = \mathbb{C}[x] / (f, J_f)$ to determine the dimensions $\mu = \dim_{\mathbb{C}} (\mathbb{C}[x] / J_f)$ and $\tau = \dim_{\mathbb{C}} \mathcal{Q}_f$. The staircase quantization is then verified by tracking the non-vanishing difference $\Delta = \mu - \tau$ across the phase diagram, confirming the locking of interaction strength to discrete integer values \cite{Tang2020}.

\section*{Acknowledgments}
This research was conducted during an independent research sabbatical in the Himalayas (Nepal). The author thanks Konstantin Dorfman (State Key Laboratory of Precision Spectroscopy, ECNU, Shanghai) for stimulating discussions regarding time crystals that inspired the initial concept of the geometric k-comb. However, the development of the underlying mathematical frameworks (DMHM, cQGT, Floquet Monodromy Spectroscopy) and the \textbf{QuMorpheus} computational implementation were executed independently by the author. The author acknowledges the global open-source community for providing the computational tools that made this work possible. LLM assistance was utilized strictly for \LaTeX\ syntax optimization and symbol consistency checks; all scientific conceptualization, derivations, and text were derived and verified by the author. The author retains the \texttt{uci.edu} correspondence address courtesy of the University of California, Irvine.

\section*{Data and Code Availability}
The core computational framework, \textbf{QuMorpheus}, used for all numerical results in this work, is open-sourced under a Copyleft license and is available at \url{https://github.com/prasoon-s/QuMorpheus} \cite{SaurabhNatComm}. Independent verification scripts (Python) are available from the author upon reasonable request. To ensure mathematical rigor, the fundamental theorems of the DMHM framework, the construction of the cQGT, and the Floquet Monodromy Spectroscopy protocol have been formalized in the \textsc{Lean 4} theorem prover; these proofs are available at \url{https://github.com/prasoon-s/LEAN-formalization-for-CMP} \cite{saurabh2025holonomic}.

\bibliography{Bibliography/bibliography_kcombs}

\end{document}